    \documentclass[preprint2]{aastex}
    \usepackage{url}
\usepackage{verbatim}
\usepackage{epstopdf}
\usepackage{graphicx}% Include figure files
\usepackage{dcolumn}% Align table columns on decimal point
\usepackage{bm}% bold math
\usepackage[english]{babel}

\begin{document}

\title{
Accretion-caused deceleration of a gravitationally \\ powerful compact stellar object moving within a dense Fermi gas
}
%% Running heads
%\shorttitle{Accretion-caused deceleration} \shortauthors{Tito \& Pavlov}

\author{E. P. Tito}
\affil{Scientific Advisory Group, Pasadena, CA 91125, USA}
\author{V. I. Pavlov}
\affil{UFR des Math\'{e}matiques Pures et Appliqu\'{e}es,
%-- LML CNRS UMR 8107,
 Universit\'{e} de Lille 1,
59655 Villeneuve d'Ascq, France
%\vspace{0.4cm}
%\textnormal{June 15, 2014}
}

\begin{abstract}
We consider accretion-caused deceleration of a gravitationally-powerful compact stellar object traveling within a cold Fermi-gas medium. We provide analytical and numerical estimates of the effect manifestation.
\end{abstract}


\maketitle

\section{Introduction} \label{S:Introduction}

Numerous fast-moving solitary stellar objects, called the "wandering stars", have been astronomically detected  
inside and outside our Galaxy.
The speeds of these objects sometimes reach as high as $1200  \, km \, s^{-1} $ \citep{g15}, 
greater than even the galactic escape velocity ($v_e \simeq 500 - 600 \, km \, s^{-1} $).
 The exact nature of these stellar bodies is uncertain, and a variety of hypotheses and formation scenarios have been proposed. 
 \citet{h88}, \citet{ree90}, \citet{knp93a}, \citet{knp93b}
Most of these scenarios involve dramatic acceleration of the object -- whether a star, a neutron star, or possibly a piece of the torn apart debris -- 
by the super-massive black hole located at the galactic core.
Indeed, the gravitational might of the black hole is such that many objects, 
even those that are deemed essentially indestructible in other circumstances, can be torn apart 
by tidal forces 
into pieces and flung out with enormous speeds. 

While a  collision of such a fast-moving object with another stellar object is, generally speaking, a low probability event, 
it is not impossible, especially when considering densely populated areas of the Galaxy and when taking a long historical perspective. 

Moreover,  a collision of a neutron star with a star -- a red giant, a supergiant, or a white dwarf --  
is proposed as one of the leading scenarios for the formation of a Thorne-Zytkow object,  
theoretically hypothesized in 1977 and potentially discovered in 2014. (\citet{tz77}, \citet{lmzm14})

While over the years much focus has been given in such scenario to various critical aspects of the phenomenon, 
to our knowledge  
the mechanism of accretion-caused deceleration of the neutron star has never been considered. 
Furthermore, this mechanism has never been considered in any scenarios of compact and expansive objects collisions.

In this paper, we specifically focus on the mechanism of deceleration resulting from the accretion of the dense surrounding medium 
onto a rapidly moving gravitationally-powerful compact object. 
We consider a generalized and intentionally simplified scenario, involving not specifically a formation of a Thorne-Zytkow object, 
but rather a head-on collision of a neutron star-like, but non-rotating and non-magnetized, compact object 
with  a dense white dwarf-like medium.

Generally speaking, in such a deceleration scenario different mechanisms can be responsible for the kinetic energy decrease of the moving object:
classical hydrodynamical drag
\citep{d64},
gravitational drag in collisionless systems
\citep{ch43} 
which is called 
dynamical friction 
in astrophysics 
\citep{o99},
Cherenkov's radiation of various waves (related to collective hydrodynamical motions) which are generated inside the medium
\citep{ps85},
\citep{pk90},
\citep{pt09},
interaction of proper magnetic field for strongly magnetized object with surrounding plasma
(see  \citet{t-a12} and Refs therein),
as well as other more complex possibilities.

Analytically, the relative importance of these various mechanisms contributing simultaneously to the aggregate deceleration, 
can be assessed using the dimension-analysis approach.

The classical hydrodynamical drag (passive resistance of the surrounding medium) for a blunt object 
moving fast enough (large Reynolds number) 
to produce a turbulent wake,
is proportional to its cross--section, i.e. $k_d  \rho R^2 V ^2 $. \citep{ll87} 
Here 
$\rho$ is the medium density, 
$R$  is the characteristic (transversal) size of the object, 
$V$ is the velocity of the object relative to the surrounding medium.   
The dimensionless drag coefficient, $k_d$, 
takes into account both  skin friction and  form factor. 
The characteristic time of deceleration due to hydrodynamical drag, $\tau_d$, is then 
$\tau_d^{-1} \sim  \rho R^2 V / M $, where $M$ is the object's mass.

Dynamic friction, called gravitational drag in astrophysics, also contributes to the loss of momentum and kinetic energy 
when a moving object gravitationally interacts with the surrounding matter (rarefied cloud). %in space.  
\citep{ch43}
The essence of the effect is that small cloud particles are pulled by gravity  toward the object, thus increasing the cloud density.  
But if the object already moved forward, the density increase actually occurs in its wake. 
Therefore, it is the gravitational attraction of the wake that pulls the object backward and slows it down.  

A simplified equation for the force from dynamical friction has the form
$M dV / dt = F_{Ch} = - C \rho G^2 M^2 / V^2$.
The dimensionless numerical factor $C$  depends on the so--called Coulomb logarithm  
and on how velocity of the object $V$ compares to the velocity dispersion $\sim \sqrt{\langle v^2 \rangle}$ 
of the cloud particles, 
i.e. on the argument $\xi = V / \sqrt{\langle v^2 \rangle}$. 
The characteristic time of the process obtained from the Chandrasekhar equation written above,
is  $ \tau_{Ch}^{-1} \sim \rho G^2 M  ( \sqrt{\langle v^2 \rangle })^{-3}$.  

For gravitationally-powerful moving objects  
that are capable of capturing the surrounding medium particles onto its surface (not just pulling them into its wake), 
when the increase of the object's mass due to the accretion is non-negligible, 
the constant-mass-body equations of motion commonly used in celestial dynamics, no longer apply. 
To our knowledge, at present there are no qualitative or numerical considerations of this effect. 
The aim of this article is to fill this void.

The characteristic time $\tau_a $ for the accretion deceleration is such that 
$\tau_a^{-1} \sim  M$: 
the more massive the body is, the faster the accretion onto it occurs. 
In this consideration, we focus on the accretion onto an object with {\em small size} but significant mass, 
and the one that moves with trans- or super-sonic speed through an non-perturbed, uniform at infinity medium that is free of self-gravity.
Then the remaining  apparent characteristics of the process are the  
gravity constant $G$, and  
density $\rho$ and pressure $P$ of the accreting medium.
The combination of these parameters that matches the dimension of (the inverse of) $\tau_a $ is 
$\tau_a^{-1} \sim \rho G^2 M (\rho / P )^{3/2}$, or $\tau_a^{-1} \sim \rho G^2 M s^{-3}$ 
where $s^2$ characterizes 
the square of speed of propagation of small density perturbations within the medium. 
(The exact analytical derivation follows in the subsequent section.) 
Despite the fact that $\tau_a^{-1} \sim  G^2$, the presence of other parameters makes $\tau_a $ range widely, 
thus indicating that the accretion effect may be negligible or dominant depending on the specific parameters at the moment.

The mathematical treatment of the deceleration process becomes significantly more complex if
proper rotation,  and/or  magnetic fields of magnetized stars, and/or  interaction with surrounding plasma 
are included. 
If the velocity, magnetic moment and angular velocity vectors 
point in different directions, 
the results are strongly dependent on the model configuration.
Magnetosphere acts as an obstacle for the incoming accreting flow, 
thus reducing the accretion rate onto magnetized objects.  
When the magnetic impact parameter $R_m$ is greater than the accretion radius $R_{ac}$ 
calculated from the classical model 
to define the region of the surrounding medium involved in the accretion process, 
accretion is not important. 
If $R_m < R_{ac}$, 
the accreted mass accumulates near magnetic poles the most. 
(See \citet{t-a12} and references therein.) 

At the dimension-analysis level, 
comparison of the characteristic time scales of all the mechanisms involved, performed for the specific circumstances of the problem at hand, 
reveals whether any of the mechanisms may be considered negligible and thus omitted. 
Obviously, the more dominant process is the one with the smaller  $\tau$.

In the following analysis, we focus exclusively on the accretion mechanism, 
and will ignore all other types of drag, magnetic and rotational effects.

\section{Accretion model}

As a physical phenomenon, accretion has been studied for a variety of settings. 
The rate of accretion for moving stars is estimated from the expression $\dot{M} = \rho V \times \pi R_{ac}^2$.  
Here $R_{ac}$ is the characteristic capture radius - the principal quantity.
The early works
(\citet{hl39} and
\citet{b52})
considered the accretion onto a stellar body moving at a constant velocity through an infinite gas nebula. Subsequently, a variety of media has been considered:
interstellar medium, a stellar wind, or a common envelope (where two stellar cores become embedded in a large gas envelope formed when one member of the binary system swells).  %by
See, among others,  
\citet{p78},
\citet{r94},
\citet{ra94},
\citet{r96},
\citet{bkp97},
\citet{p00},
\citet{ts00},
\citet{b01},
\citet{ec04},
\citet{t-a12}.
Accretion onto a neutron star from the supernova ejecta has also been extensively researched -- for a radially-outflowing ejecta
\citep{c71},
\citep{zin72},
for an in-falling ejecta
\citep{c89},
\citep{csw96}
and when the object is moving at a high speed across the supernova ejecta
\citep{z-a07}.

These prior studies have considered media with low or moderate density.  
In this article, we provide an analysis for high density medium, 
such as the degenerate dense Fermi gas, examples of which are 
white or black dwarfs. 
These dwarfs are the final stages in the evolution of stars not massive enough ($M<9 \, M_{\odot}$) to collapse into a neutron star or undergo a Type II supernova.  
They are composed of electron-degenerate matter with densities exceeding $10^7 \, kg/m^3$. 
A black dwarf is a white dwarf that has sufficiently cooled to no longer emit visible light.

{\bf Equation of motion for body of variable mass.} The equation of motion for a body of variable mass follows from the law of conservation of linear momentum of the {\it entire} system  composed of the object and the surrounding mass captured by the object.
Thus, when an object enters a dense gaseous "cloud", and surrounding nebula particles accrete onto the gravitationally powerful object, the motion of the object will be described by
$ \Delta ( M_* {\mathbf V} ) - {\mathbf v} \, \Delta M_* = \Delta {\mathbf I}$
\citep{m97}.
Here $M_* (t)$ and  ${\mathbf V} (t)$  denote, respectively, the mass and  velocity of the moving object in an inertial frame at instance $t$, ${\mathbf v}(t)$ is the velocity (in the same frame) of the accreting nebula particles which compose mass $\Delta M_*$), and $\Delta (...)$ denotes change of quantities over the small finite interval of time $\Delta t$. Qualitatively, this is the simplest model when particles of environment "stick" to the "attractor". Quantity $\Delta {\mathbf I} = M_* {\mathbf w} \, \Delta t$  is the impulse of an external force ${\mathbf F} = M_* \, {\mathbf w}$. Here, ${\mathbf w}(t)$ is acceleration of the object in an inertial frame. Then it follows (in form of increments):
\begin{equation}
\label{m1b}
\Delta {\mathbf V} + ({\mathbf V} - {\mathbf v}) \, {\Delta M_*} / {M_*}  = {\mathbf w} \Delta t.
\end{equation}
If the increment $\Delta M_* \rightarrow 0$, we obtain the classical Newtonian equation of motion for bodies of fixed mass. When the object mass changes, ${\Delta M_*} \neq 0$, the concept of a "steady-moving" body in absence of external actions (${\mathbf w} = 0$) is not a precise one. Eq.~\ref{m1b} is the basis of equations describing the rocket motion. The elementary work of the "accretion" force which is proportional to $\Delta M_*$, is
$    \Delta {A} = - ( {1}/{2} ) {\mathbf v}^2 \,  \Delta {M}$.  This work is negative when $\Delta {M} > 0$ and therefore, the reduction of kinetic energy takes place (deceleration occurs). Obviously, this expression must be statistically averaged with respect to all possible values of velocities ${\mathbf v}$ of the accreting particles  for the given $\Delta M$ (see the main text of the paper). The part of this work is transformed into heat received by the object.
This quantity (per unit time) can be estimated as $\dot{Q} \simeq \dot{M} \langle {\mathbf v} \rangle^2 $ (to within a factor of the unit order).

Eq.~(\ref{m1b}) must be statistically averaged with respect to all possible values of velocities ${\mathbf v}$ of the accreting particles  for the given $\Delta M_*$. After the averaging, the velocity  ${\mathbf v}$ of accreting fragment $\Delta M_*$ in Eq.~(\ref{m1b}) which contains a large number of accreting particles is replaced by averaged $\langle {\mathbf v} \rangle$, and transition $\Delta t \rightarrow dt$ is performed to write Eq.~(\ref{m1b}) in terms of derivatives.

{\bf Calculation of averaged velocity of accreting particles.} The following step is to find the expression for $\langle \mathbf{v} \rangle$  which obviously is not zero for the moving body in accordance with the simple philosophy that the body will collide with particles flying in face more often than with particles that are catching up him.

We assume the spherical symmetry of the velocity distribution of the gas particles and their spatial homogeneity, so that distribution function $f ({\mathbf v}) = f (v)$ is a function of velocity module. The probability that any gas particle occupies element $d w = dv_x dv_y d v_z$ in the space of velocities is proportional to $d w \, f (v).$ The probability of the object to capture the gas particle with velocity ${\mathbf v}$ is proportional to the cross--section of interaction, i.e. to the product of the module of relative velocity of the particle with respect to the object ($|{\mathbf v} - {\mathbf V}|$) and $d w \, f ( v )$. Thus, the average velocity is
\begin{eqnarray}
\label{vm1}
\langle {\mathbf v} \rangle =
{\int d w \, {\mathbf v} \, |{\mathbf v} - {\mathbf V}| \, f (v)} \bigg/ {\int d w \, |{\mathbf v} - {\mathbf V}| \, f (v)}
\end{eqnarray}
Due to the axial symmetry of the problem, $\langle {\mathbf v} \rangle$ is co-linear with ${\mathbf V}.$ In the spherical coordinate system with
$d w = $ $2 \pi d \theta \, \sin \theta \, d v \, v^2$ where $\theta$ is the angle between ${\mathbf v}$ and ${\mathbf V},$
\begin{eqnarray*}
\langle v \rangle = \nonumber \\
\frac{ 2 \pi \int_0^{\infty} \int_0^{\pi} d v \, v^2 \, d \theta \, \sin \theta \, f (v) v \, \cos
\theta \, \sqrt{v^2 + V^2 - 2 v V \cos \theta} }{ 2 \pi \int_0^{\infty} \int_0^{\pi} d v \, v^2 \, d
\theta \, \sin \theta \, f (v) \, \sqrt{v^2 + V^2 - 2 v V \cos \theta} } ,
\end{eqnarray*}
which, after integrating with respect to angle $\theta, $ produces the following expression:
\begin{eqnarray*}
\langle v \rangle = \nonumber \\
\frac{\int_0^V d v \, v^3 f (v) [\frac{2}{3} v - \frac{2}{15}
\frac{v^3}{V^2}] + \int_V^{\infty} d v \, v^3 f (v) [\frac{2}{3} V - \frac{2}{15}
\frac{V^3}{v^2}]}{\int_0^V d v \, v^2 f (v) [2 V + \frac{2}{3} \frac{v^2}{V}] + \int_V^{\infty} d v \,
v^2 f (v) [2 v + \frac{2}{3} \frac{V^3}{v^2}]} .
\end{eqnarray*}

\begin{figure}[h!]
\centering
\includegraphics[width=7.0cm]{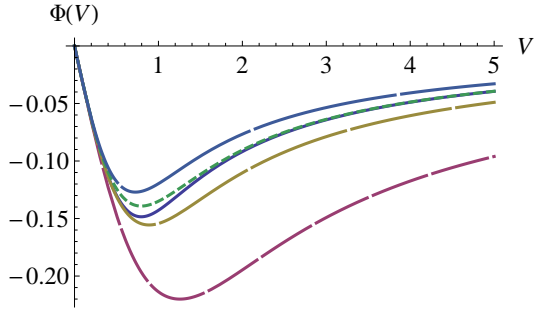}
\caption{
Velocity correction $\Phi (V) \sim - \langle v \rangle$ calculated as a functional of different functions of distribution:
the Fermi distribution (where function $f (v)$ is proportional to the Heaviside function, $f (v) \sim H (v_F - v)$) (base line),
and the Maxwell distribution (where $f (v) \sim \exp (- v^2 / \beta v_{Fp}^2)$) shown for $\beta = 1, \, 2, \, 2.5, \, 3$ (lower to upper lines).
In the presented example, both Fermi and Maxwell distributions are similar when $\beta = 2.5$.
}
\label{averv}
\end{figure}

The obtained expression permits the use of any distribution function, both the Maxwell-Boltzmann $f (v) \sim \exp (- v^2 m / 2 k T)$ for high temperatures and the Fermi one for low temperatures. Technically, both distributions give similar results (Fig.~\ref{averv}).

We assume that the surrounding gas is composed purely of ionized hydrogen--degenerate electron--proton plasma. Since $m_p \gg m_e$, only the proton component is significant for the object mass change.

We consider in more detail  the distribution for the full degenerate Fermi gas of proton component  which is valid when the temperature of medium $T \ll T_{Fp}$.
For the distribution with respect to velocities of fully degenerate non--relativistic Fermi gas of protons/nuclei
\citep{f62},
\citep{f29},
$f (v) \sim H ( v_{Fp} - v)$, where $v_{Fp}$ is the local Fermi boundary velocity of the nebula heavy particles and $H ( \xi )$ is the Heaviside step function. Parameter $v_{Fp}$ is defined as $v_{Fp} = (6 \pi^2 /2)^{1/3} (\hbar / m_p ) (\rho / m_p )^{1/3} = (m_e / m_p) v_{Fe} \ll v_{Fe}$. Numerically this gives $v_{Fp} = 1.643 \times 10^2 \rho^{1/3} \; m \, s^{-1}$ where $\rho$ is measured in $kg \, m^{-3}$. Even for rather large densities of accreting medium (for example, for a white dwarf near the boundary of stability $\rho = 10^9 \, kg \, m^{-3}$) when $v_{Fe}$ becomes relativistic, for proton component $v_{Fp}  \ll c = 2.99 \times 10^8 \, m \, s^{-1}$. When temperatures of $p-$ and $e$--components of the medium are of the same order, parameter $v_{Fp}$ is of the same order as the speed of sound in the medium. To simplify the subsequent analysis, we introduce dimensionless velocity, $V / v_{Fp} \rightarrow V$, and
express $\langle v \rangle$ as $\langle v \rangle \equiv  v_F \Phi (V)$ with function
\begin{eqnarray}
\Phi (V) =  \frac{-4 + 28 V^2 + \alpha_1(V) H(1 - V)}{7 V (-4 (1 + 5 V^2) + \alpha_2 (V) H(1 - V))} < 0.
\quad \label{v-m2}
\end{eqnarray}
Here, $\alpha_1 (V)=(-1 + V)^4 (4 + 16 V + 12 V^2 + 3 V^3)$, $\alpha_2 (V) = (-1 + V)^4 (4 + V)$ and $H (s)$ is the Heaviside function.
Behavior of $\Phi (V)$  is given in Fig.~\ref{averv}.
A good polynomial approximation of $\Phi (V)$ is
$
\Phi (V) \simeq  (- 0.333 V  + 0.318 V^3 - 0.135 V^5 )H(0.795 - V) + \\
( - 0.2 V^{-1}
+ 0.067 V^{-3} - 0.009V^{-5}) H(V - 0.785).
$

{\bf Mass accumulation.} Then Eq.~(\ref{m1b})  takes form suitable for our analysis:
$\label{eqm1b}
{d {V}} / {(V - \Phi (V))} =  - {d M_*} /{M_*} $.
The mass and the time-derivative of the mass of the object are expressed in terms of $V$  as
\begin{eqnarray}
\label{eqm1e1}
M_* (t) = M_0 \exp \int_{V(t)}^{V (0)} \frac{d \xi}{\xi - \Phi (\xi)}
\equiv M_0 \exp J(t) .
\end{eqnarray}
Further, we transition to dimensionless variables and express the object mass $M_*(t)$ in terms of (constant) $M_0$ (which from this point on will be called the \emph{initial} mass of the object) and the normalized variable  $m(t)$, so that  $M_*(t) = M_0 m(t)$. We assume for simplicity the density of the accreting medium to be constant $\rho_F$.
The dimensionless mass of the object evolves as
\begin{eqnarray}
\label{eqm1e2}
 \dot{m} =  - (V - \Phi (V)^{-1} \exp J (t) \dot{V} .
\end{eqnarray}
showing that when $\dot{V}  < 0$, deceleration, mass $m (t)$ is increasing, $\dot{m}  > 0$.

{\bf Regimes of motion and results of calculation.}
To close  the system of the equations, we have to propose an evolution equation for the mass of the object,  
i.e. $\dot{M}_* = \dot{M}_* (M_*, V, ...)$.

The simplest model is to assume that the object %perturber 
mass increases due to the simple "adhesion" of the surrounding particles and  that its mass increases proportionally to the effective surface area, 
i.e.  $\dot{M}_* \sim 4 \pi R^2 (t)$ (Appendix~\ref{ap:2}).

A more complex model includes  
the traditional interpolation  for $ \dot{M}_*$ (proposed in  \citet{b52}  for accretion onto both  a resting and  a moving object)
(see also,
\citet{st83}, p.~420).
Bondi--Hoyle--Lyttleton (BHL) accretion, in its simplest form, considers a point mass moving through a gas cloud 
that is presumed to be non-self-gravitating and uniform at infinity.
Gravity focuses the gas cloud particles behind the point mass. 
Gas particles then accrete to the mass.
%In its purest form, Bondi--Hoyle--Lyttleton (BHL) accretion concerns the motion of a point mass through a gas cloud. 
%The cloud is assumed to be free of self--gravity, and to be uniform at infinity. 
%Gravity focuses material behind the point mass, which can then accrete some of the gas. 
%This problem has found applications in many areas of astrophysics 
(See \citet{ec04} and Refs therein.)
This expression can be presented (with a small reformulation) in form
\begin{eqnarray}
\frac{\dot{M}_* }{M_*} = 4 \pi \kappa G^2  M_*  \frac{ \rho^{5/2} }{ (\rho V^2 + P)^{3/2}}
\label{b0f}
\end{eqnarray}
Here, the left part of equation represents the rate of mass change of the object,  %perturber, 
the right one is defined by factors which governs the accretion process, $\rho$ is a characteristic medium density, $V$ is the velocity of the object with respect to medium, $s$ is the (isothermical) sound speed in the medium  (at large distance from the object), $P = \rho s^2$, numerical coefficient $\kappa$ is of the order of unity. 
(See Appendix~B.)

The BHL formula written in the form of Eq.~(\ref{b0f}) shows that it can be obtained from simple arguments based on the dimensional analysis. In fact, the rate of accretion ${\dot{M}_* } / {M_*}$ has to be faster when the object %perturber 
is massive, i.e. ${\dot{M}_* } / {M_*} \sim {M_*}$. The process is governed by the gravity, $G$, and by the principal properties of the medium: density, $\rho$, and pressure $P (\rho, ...)$ which determines the equation of state. The only dimensional combination of $G$, $\rho$ and $P$ which produces the necessary dimension, is $M_* \times G^2 \rho^{5/2} P^{-3/2} $. In a moving medium, the pressure must be replaced by the dynamical pressure $P + \rho V^2$. From here, Eq.~(\ref{b0f})  is obtained.

In such case (and with $M_* \rightarrow M_0 m (t)$ , $P = \rho s^2$  and $V \rightarrow v_{Fp} V$) the dimensionless form of the equation takes a simple form.
By combining Eq.~(\ref{b0f}) with Eqs.~(\ref{m1b}), (\ref{eqm1e1}) and (\ref{eqm1e2}), we obtain
\begin{eqnarray}
\frac{d V}{d t} = - \bigg( 4 \pi \kappa \frac{\rho G^2 M_0}{s^3}  \bigg) \frac{V - \Phi (V)}{(\epsilon V^2 + 1)^{3/2}} \, \exp (-J (t)) .
\label{b1f}
\end{eqnarray}
Here $\epsilon = m_e / m_p \ll 1$. Coefficient $\beta$  (expression in parentheses)
can be written as  $\beta \equiv \tau^{-1} = (4 \pi \kappa ) \rho G^2 M_0 /s^3 = 1.995 \times 10^3 (4 \pi \kappa ) (M_0 / M_{\odot})$,
where
$M_{\odot} = 1.989 \times 10^{30} \, kg$ is the Sun mass, $\tau$ is the characteristic time scale. Function $J (t)$ is defined by Eq.~(\ref{eqm1e1}). Eqs.~(\ref{v-m2})--(\ref{b1f})  complete the system of necessary equations.  Eq.~(\ref{b1f}) establishes the time scale $\tau^{-1} = (4 \pi \kappa) (\rho G^2 M_0 / s^3) $, which depends on the initial mass of the object and the properties of the target medium. By expressing the physical parameters of the problem in units $\tau$ (for time) and $v_{Fp}$ (for velocity), we obtain the universal solution for the basic set of equations.

\begin{figure}[h!]
\centering
\includegraphics[width=7.0cm]{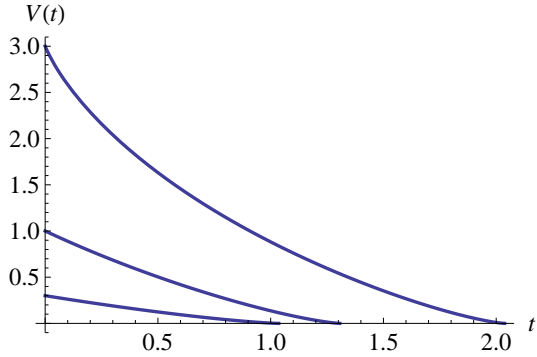}
\caption{
Evolution of the colliding object's velocity,
for initial velocities $V_0 = 0.3$ (lower line),  $V_0 = 1$ and $V_0 = 3$ (upper line).
}
\label{velocities}
\end{figure}

\begin{figure}[h!]
\centering
\includegraphics[width=7.0cm]{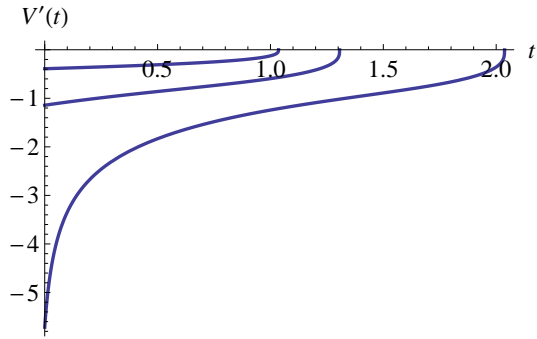}
\caption{
Evolution of the colliding object's (dimensionless) deceleration,
for initial velocities $V_0 = 3$ (lower line),  $V_0 = 1$ and $V_0 = 0.3$ (upper line).
}
\label{decelerations}
\end{figure}

\begin{figure}[h!]
\centering
\includegraphics[width=7.0cm]{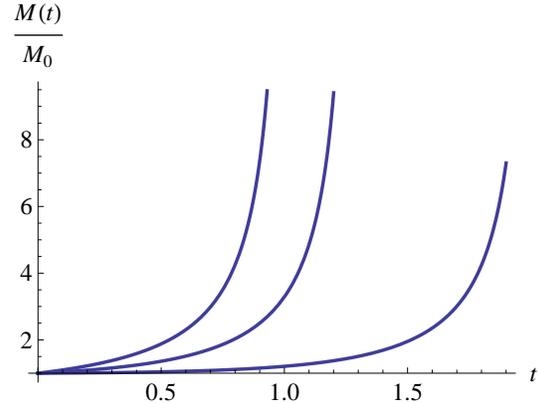}
\caption{
Normalized (by $M_0$) total mass of the colliding object as function of time for different initial velocities: $V_0 = 3, \; 1, \; 0.3$ (from lower line to upper line).
When the object stops, $V \rightarrow 0$, the accreted mass $M \rightarrow \infty$.
In such situation, the used approximations are no longer valid, and the general relativity approximation has to be taken into consideration.
}
\label{mass}
\end{figure}

\begin{figure}[h!]
\centering
\includegraphics[width=7.0cm]{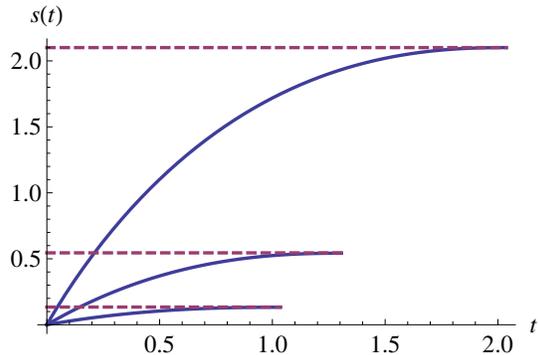}
\caption{
The dimensionless path of
the object
as function of time for different initial velocities: $V_0 = 0.3, \; 1, \; 3$ (from lower line to upper line).
}
\label{distances}
\end{figure}

The greater is the density of the surrounding medium, the stronger is the deceleration effect.

Figs~\ref{velocities}--\ref{distances} illustrate the model results for several initial conditions.
The figures show the evolution of each scenario (characterized by the initial dimensionless velocities $V_0 = 0.3, \; 1, \; 3$) until the object decelerates to a stop. Note that the state of full stop is an asymptotic state. Neither the Bondi formula, nor the modified version of it presented in this analysis, properly describe the process in the vicinity of such state. It implies that the entire (infinite) mass of the accreted medium is captured by the object within the finite period of time. In reality, of course, the target has a finite mass, and once it is captured no further accretion (and therefore deceleration) occurs.

However, in the domain of parameters where the model is reasonably accurate and valid, it appears that in all three scenarios the object significantly decelerates (approaches its full stop) once it accretes the amount of mass equal to several times its initial mass. Which means that if the relative sizes of the object and the target are such that the entire accreted mass is "small" (not sufficient to decelerate the object to a full stop), the deceleration would vanish once the entire target mass is accreted or once the object exits the zone of influence.

Fig.~\ref{decelerations} also reveals that the magnitude of deceleration is the greatest at the beginning of the process and can be quite non-negligible.

Specific evolution scenarios depend on the %mutual 
relationship between several characteristic times: time-of-flight $\tau_p \simeq R/V_0$, or $\simeq (\rho_{object}/\rho_{target})^{1/3}R/V_0$, time of hydrodynamical �unloading� $\tau_h \simeq R/s$ inside of target, and characteristic time of accretion $\tau_a$. Here, $R$ is the characteristic size of the target. 
Depending on the combination of these parameters, different outcomes occur affecting the states of the involved bodies. However, the comprehensive analysis of such scenarios is beyond the scope of this publication.

\section{Conclusion}

Accretion-caused deceleration occurs when a gravitationally-powerful object moves through a medium, captures the surrounding particles, and decreases its kinetic energy and momentum as its mass increases. 
In this article, we presented an analysis of such scenario for 
a compact (small in size), 
non-rotating and non-magnetized,  
gravitationally-powerful object 
colliding head-on (simple model geometry) 
with 
a high-density medium (a white or black dwarf, for example). 
By describing the motion of the variable-mass body,
we demonstrated that the magnitude of the deceleration 
(caused only by accretion and no other mechanisms) 
may indeed be substantial depending on the initial conditions. 
There are several implications stemming from this result.

First, 
as mentioned earlier, 
one of the hypothesized scenarios for the formation of a Thorne-Zytkow object 
(a red giant or supergiant containing a neutron star at its core) 
is a collision of the two objects, the giant and the neutron star. 
In our demonstration, despite its intentional simplicity, 
the results at the qualitative level 
appear to be consistent with such scenario.  
As shown in Fig.~\ref{distances}, 
the neutron star may be completely captured by the target (the full-stop case), 
or the neutron star may accrete mass from the target without stopping.  
In both cases, 
the resulting object may be described as a neutron star surrounded by a gigantic envelope.
The exact outcome would depend on the  initial characteristics of the involved objects,
producing a TZO with   
a larger-sized giant in the first case and a TZO with a smaller-sized giant in the second.
Therefore, 
accretion may play an important role in the formation of the Thorne-Zytkow objects, 
even if taken as a stand-alone mechanism, 
and thus its contribution should not be neglected in  complex and more realistic multi-mechanism models. 

In this article, we derived the proper mathematical description 
for the highly dense medium (degenerate Fermi-gas) 
that is better suited for targets like white or black dwarfs, whose densities exceed $10^7 \, kg/m^3$. 
Prior studies of accretion considered only low or medium density media.

Second, 
while the accretion-caused deceleration effect is interesting on its own, 
when it is applied to the stellar objects composed of nuclear matter  
with particular equations of state (EOS),  
the situation deserves a special attention.
 As well known, 
just like  traditional matter, 
 nuclear matter  
 has its critical state with its critical temperature and density. 
(See, for example,  \citet{j83}, \citet{j84}, \citet{apr98}, \citet{k06} and 
\citet{k-a11} 
and references therein.) 
This means that if the matter of the elastic stellar object is in the state close to the boundary of liquid/gas phase transition 
(near the spinodal zone where the matter can transition into the state of "nuclear fog"), 
then speed of density perturbation propagation is close to zero. 
Then even relatively small deceleration may lead to strong stratification of the interior matter of the compact object. 
The space scale of this stratification is defined by the ratio of sound speed square and deceleration magnitude.    
Zones of compression and decompression appear throughout the compact object interior.   
Within the decompression zones, 
in the environment of the nuclear fog, 
explosive nuclear reactions (fusion and fission of fragments) may start. 
(\cite{tp13}  examine this in more detail.)

To conclude, while accretion onto neutron stars (and other compact gravitationally-powerful stellar objects 
such as fragments of a neutron star, 
quark star, strange stars, etc.)  
would rarely occur as a stand-alone process, 
in some cases 
it may meaningfully contribute to the aggregate deceleration experienced by the stellar objects. 
For non/low-magnetized objects, accretion may actually play the dominant role in the deceleration 
of the objects when they collide with other stellar bodies or traverse an encountered medium. 
In this article, we provided a new model for the treatment of dense accreting medium.

\appendix

\section{Deceleration due to adhesion}
\label{ap:2}

Consider a body moving in a dense medium. We suppose that the interaction of the surrounding particles with the body is governed by short-range  forces. Such interaction is modeled by the adhesion mechanism where particles of the environment simply adhere to the body.  Consequently, the mass and the volume of the moving body increase.  We suppose that the rate of mass increase is proportional to the surface of the body and density of environment., i.e. $\dot{M} (t) \sim \rho_e \, 4 \pi r^2 (t)$. Here, $r (t)$ is the radius of the body, $\rho_e$  is the density of the medium  (a classical example is of the  drop which is moving in a saturated vapor of water).

We suppose that the density of the body stays constant during the process at least in leading approximation.  The mass is $M = \rho_b (4 \pi /3) r^3 $. From $d / dt [ \rho_b (4 \pi /3)  r^3 ]= v_* \rho_e \, 4 \pi r^2 $, we can find  that the rate of radius increase is constant $\dot{r} = v_* (\rho_e / \rho_b)$.  Parameter $v_*$ has the dimension of velocity.  The meaning of this parameter is the characteristic velocity of adhesion of particles of the environment to the surface of the body. This parameter is determined by the regime of plasma-dynamical flow in the surrounding medium which is not a trivial problem because the process of adhesion depends strongly on the model of the environment,  for example on the equation of state of the surrounding matter.

The classical equation of motion of the body of variable mass in presence of the traditional hydrodynamical  drag is  $M \dot{\mathbf v} = - C \rho_e r^2 v^2  {\mathbf n}_v +  {\mathbf c} \dot{M}$,  where ${\mathbf c} = {\mathbf V} - {\mathbf v}$. Here, $\mathbf V$ is the velocity of the medium particle adhered to the body in an inertial frame, $\mathbf v$   is the body velocity in the same frame, the drag is proportional to square of the body velocity, dimensionless parameter $C$ is of order unity.

We suppose for simplicity that all particles of the environment are immobile in the initial non--perturbed state, ${\mathbf V} = 0$. This is assumed to simplify the consideration and to obtain an analytical solution.  The equation of motion becomes
\begin{equation}
M \frac{d v}{dt} = - C \rho_e r^2 v^2  - v \frac{d M}{dt} ,  \quad  \rightarrow \quad
%\end{equation}
%or
%\begin{equation}
\frac{d }{dt} (\frac{4 \pi}{3} r^3 v) =   - C \frac{\rho_e}{\rho_b} r^2 v^2.
\end{equation}
Since $(d / d t) ... = (d r / dt) (d / dr) ... = v_* (\rho_e / \rho_b) (d / dr) ...$ and $\xi = r^3 v$, we obtain the simple equation
\begin{equation}
\frac{d \xi }{\xi^2} = - ( \frac{3 C}{4 \pi v_*} )  \frac{d r}{r^4}
\end{equation}
which can be resolved analytically :

\begin{equation}
v (t) = \frac{v_0}{s^3 + \alpha  v_0  ( s^3 -  1 )}  .
\label{sol}
\end{equation}
Here, argument  $s = (r_0 + v_* (\rho_e / \rho_b) t) / r_0$,  $\alpha = C / 4 \pi v_*$ and  $r = r_0$ and $v = v_0$ at $t =0$.  It follows from here that the characteristic time of the process of deceleration is $\tau \sim r_0 / v_* (\rho_b / \rho_e) $. For $t \gg \tau$ and $v \gg v_*$, the regime of deceleration becomes the universal one and independent on an initial velocity of the body. Obviously, Eq.~(\ref{sol})  should be regarded only as the zero--approximation in the averaged on all possible values of parameter $\langle {\mathbf V} \rangle$, which is obviously not zero  for the moving body in accordance with the simple observation that the body will collide with the particles flying in its face more often than with the particles that are catching up to it.

The deceleration can be written now as
\begin{eqnarray}
a = - \frac{3 C}{4 \pi} \frac{\rho_e}{\rho_b} \frac{v^2}{r}  - 3  \frac{ \rho_e}{\rho_b} \frac{v_* v}{r},
\end{eqnarray}
or
\begin{eqnarray}
a = - \frac{3 C}{4 \pi} ( \frac{\rho_e}{\rho_b} ) \frac{v_0^2}{r_0}  \frac{1}{s (s^3 + \alpha  v_0  ( s^3 -  1 ))^2}  -
%\nonumber\\
3 (\frac{\rho_e}{\rho_b}) \frac{v_* v_0}{r_0} \frac{1}{s (s^3 + \alpha  v_0  ( s^3 -  1 ))} ,
\end{eqnarray}

If parameter $v_* \sim v_0$, the mechanism of deceleration due to adhesion may be comparable in magnitude with the mechanism of deceleration due to drag.

\section{Interpolating expression for pressure in medium.}

To assess  the form of the EoS of the target medium, we consider here the simplest plasma composed from protons
 and electrons.

The form of equation of state depends on how the temperature of the medium $T$ compares to characteristic temperatures.
One can introduce the following characteristic temperature parameters:
temperature of ionization $T_i \sim 10 \, ev$,
Fermi temperature for proton component $T_{Fp} (\rho )$,
Fermi temperature for electron component $T_{Fe} (\rho ) \sim (m_p / m_e) T_{F p} \sim 10^{3} T_{F p}$ for non-relativistic electrically neutral plasma,
and temperature $T_r \sim m_e c^2 \simeq 0.5 \, Mev$ when relativistic effects must be taken into consideration.
Obviously,  $T_i \ll T_{Fp} \ll T_{Fe} \ll T_r$. We consider the case when  $T > T_i$.

We assume for simplicity  that temperatures of the two components of plasma are of the same order: $T_e \simeq T_p = T$. Electrical neutrality of plasma signifies  $n_p = n_e =n$.  Here, $n_{e,p}$ is the number of free electrons/protons per unit volume. Masses of protons and electrons satisfy $m_p \gg m_e$. The density of plasma is $\rho = m_p n_p + m_e n_e \simeq m_p n$. The form of pressure depends on the level of temperature $T$ relative to the Fermi temperatures $T_{Fp}$ and $T_{Fe}$.

For high  temperatures, $T > T_{Fe} = E_{Fe} $ where $E_{Fe}$ is the Fermi energy for electron component,
 both components of the plasma can be considered as classical gas, i.e. the equation of state is $P =  T( n_p + n_e)  \simeq ( {2  T }/{m_p}) \rho $. For low  temperatures, $T < T_{Fp}$, the pressure is essentially determined by the degenerate electron component, because of $m_p \gg m_e$, for which the electron degeneracy pressure in a medium can be computed as
\begin{eqnarray*}
P = \frac{2}{3} \frac{E_{tot}}{V} =
\frac{(3 \pi^2)^{2/3} \hbar^2}{5 m_e} n_e^{5/3} \equiv  \frac{2}{5} n E_F= \frac{(3 \pi^2)^{2/3} \hbar^2}{5 m_e m_p^{5/3}} \rho^{5/3} .
\end{eqnarray*}
Here, $\hbar $ is the reduced Planck constant.

When electron energies reach relativistic levels (white dwarf with mass $0.3  M_{\odot}< M <  M_{Ch}$; the Chandrasekhar limit $ M_{Ch} \sim 1.4 M_{\odot}$), a modified formula is required, $P \sim \rho^{4/3}$: for the relativistic degenerated matter, the equation of state is "softer" and $E_{Fe} = \hbar c (3 \pi^2 n_e)^{1/3} $ for the ultra--relativistic case. In fact, pressure scales with density as $n^{5/3}$ provided that the electrons remain non--relativistic (speeds $v \ll c$). This approximation breaks down when the white dwarf mass is close to the boundary of stability to become a neutron star. The relativistic and non--relativistic expressions for electron degeneracy pressure are approximately equal at about
$n_e =10^{36} \, m^{-3}$, about that of the core of a $0.3 \, M_{\odot}$ white dwarf. As long as the star is not too massive, the Fermi pressure prevents it from collapsing under gravity and becoming a black hole.

So, we can use the simple interpolation expression for qualitative estimation of EoS in domain $T_{Fp} < T < T_{Fe}$ when one can neglect the relativistic effects:
\begin{eqnarray}
P \simeq \frac{2 T }{m_p} \rho + \frac{(3 \pi^2)^{2/3} \hbar^2}{5 m_e m_p^{5/3}} \rho^{5/3}
\end{eqnarray}
Quantity  $P / \rho$ has the dimension of the square of velocity 
and determines the order of  square of speed of propagation of small density perturbations in a medium.

In domain $T < T_{Fp}$, both components of plasma are degenerated, and we obtain
\begin{eqnarray}
P \simeq  \frac{(3 \pi^2)^{2/3} \hbar^2}{5 m_p^{5/3}} (\frac{1}{m_p} + \frac{1}{m_e} ) \rho^{5/3}
 \simeq \frac{(3 \pi^2)^{2/3} \hbar^2}{5 m_e m_p^{5/3}} \rho^{5/3}
\end{eqnarray}

\section{Conflict of Interests}

The authors declare that there is no conflict of interests regarding the publication of this paper.

%\section{Acknowledgments}

%The authors would like to thank the [] for the financial assistance for this work  and the [] for providing all facilities.

\end{document}